\documentclass[prl,letterpaper,aps,floatfix,superscriptaddress,twocolumn]{revtex4-1}
\usepackage{graphicx}
\usepackage{amsmath}
\usepackage{amsfonts}
\usepackage[small,bf]{subfigure}
\usepackage{enumitem}
\usepackage{color}
\newcommand{\beq}{\begin{equation}}
\newcommand{\eeq}{\end{equation}}
\newcommand{\bk}{{{\bf{k}}}}

\newcommand{\bQ}{{{\bf{Q}}}}

\newcommand{\bG}{{{\bf{G}}}}

\newcommand{\bB}{{\bf{B}}}

\newcommand{\beqa}{\begin{eqnarray}}
\newcommand{\eeqa}{\end{eqnarray}}
\newcommand{\pdg}{{\vphantom \dag}}
\newcommand{\dg}{{\dag}}
 
\newcommand{\bsigma}{{\boldsymbol \sigma}}

\newcommand{\upa}{\uparrow}
\newcommand{\da}{\downarrow}

\begin{document}
\title{Fractional Quantum Hall Effect in Weyl Semimetals}
\author{Chong Wang}
\affiliation{Perimeter Institute for Theoretical Physics, Waterloo, Ontario N2L 2Y5, Canada}
\author{L. Gioia}
\affiliation{Department of Physics and Astronomy, University of Waterloo, Waterloo, Ontario 
N2L 3G1, Canada} 
\affiliation{Perimeter Institute for Theoretical Physics, Waterloo, Ontario N2L 2Y5, Canada}
\author{A.A. Burkov}
\affiliation{Department of Physics and Astronomy, University of Waterloo, Waterloo, Ontario 
N2L 3G1, Canada} 
\date{\today}
\begin{abstract}
Weyl semimetal may be thought of as a gapless topological phase protected by the chiral anomaly, where the symmetries involved in the anomaly are the $U(1)$ charge conservation and the crystal translational symmetry. 
The absence of a band gap in a weakly-interacting Weyl semimetal is mandated by the electronic structure topology and is guaranteed as long as the symmetries and the anomaly are intact. 
The nontrivial topology also manifests in the Fermi arc surface states and topological response, in particular taking the form of an anomalous Hall effect in magnetic Weyl semimetals, whose magnitude is only determined by the location of the Weyl nodes in the Brillouin zone. 
Here we consider the situation when the interactions are not weak and ask whether it is possible to open a gap in a magnetic Weyl semimetal while 
preserving its nontrivial electronic structure topology along with the translational and the charge conservation symmetries. 
Surprisingly, the answer turns out to be yes. The resulting topologically ordered state provides a nontrivial realization of the fractional quantum Hall effect in three spatial dimensions in the 
absence of an external magnetic field, which cannot be viewed as a stack of two dimensional states. Our state contains loop excitations with nontrivial braiding statistics when linked with lattice dislocations. 
\end{abstract}
\maketitle
Weyl semimetal is the first example of a bulk gapless topological phase~\cite{Weyl_RMP,Burkov_ARCMP,Felser_ARCMP,Hasan_ARCMP}. 
The gaplessness of the bulk electronic structure in Weyl semimetals is mandated by topology: there exist closed surfaces in momentum space, which carry nonzero Chern numbers (flux of Berry curvature through the surface), which makes the presence of a band-touching point inside the Brillouin zone (BZ) volume, enclosed by the surface, inevitable. 
This picture, however, relies on separation between the individual Weyl nodes in momentum space, which involves symmetry considerations. 
In particular, either inversion or time reversal (TR) symmetry need to be violated in order for the Weyl nodes to be separated. 
In addition, crystal translational symmetry needs to be present, since otherwise even separated Weyl nodes may be hybridized and gapped out. 

A very useful viewpoint on topology-mandated gaplessness is provided by the concept of quantum anomalies. 
The best known example of this is the gapless surface states of three dimensional (3D) TR-invariant topological insulator (TI). 
The relevant anomaly in this case is the parity anomaly: the $\theta$-term topological response of the bulk 3D TI~\cite{Qi08} 
violates TR (and parity) when evaluated in a sample with a boundary. 
This anomaly of the bulk response must be cancelled by the corresponding anomaly of the gapless surface state~\cite{Rosenow13}, which is simply the parity anomaly 
of the massless 2D Dirac fermion~\cite{Semenoff84,Redlich84,wittenreview}.

Analogously, the gaplessness of the bulk spectrum in Weyl semimetals may be related to the chiral anomaly~\cite{Adler69,Jackiw69}. 
Suppose we have a magnetic Weyl semimetal with two band-touching nodes, located at $\bk_{\pm} = \pm \bQ = \pm Q \hat z$. 
Crystal translations in the $z$-direction act on the low-energy modes near the Weyl points as chiral rotations
\beq
\label{eq:1}
T_z^\dg c^\dg_{\pm \bQ} T_z^\pdg = e^{\mp i Q} c^\dg_{\pm \bQ}, 
\eeq
where we have taken the lattice constant to be equal to unity (we will also use $\hbar = c = e = 1$ units throughout the paper). 
However, the chiral symmetry of Eq.~\eqref{eq:1} is anomalous: an attempt to gauge this symmetry fails and produces a topological term~\cite{Zyuzin12-1}
\beq
\label{eq:2}
S = - \frac{1}{4 \pi^2} \int d t \, d^3 r \,Q_{\mu} \epsilon^{\mu \nu \alpha \beta} A_{\nu} \partial_{\alpha} A_{\beta}, 
\eeq
which expresses the impossibility to conserve the chiral charge and underlies all of the interesting observable properties of Weyl semimetals. 
In particular, variation of Eq.~\eqref{eq:2} with respect to the electromagnetic gauge potential gives the anomalous Hall conductivity 
of the Weyl semimetal 
\beq
\label{eq:3}
\sigma_{xy} = \frac{1}{2 \pi} \frac{2 Q}{2 \pi}, 
\eeq
which depends only on the separation $2 Q$ between the Weyl nodes in momentum space. 
By Wiedemann-Franz law, Eq.~\eqref{eq:3} also implies a thermal Hall conductivity
\beq
\label{eq:4}
\kappa_{xy} = \sigma_{xy}\left(\frac{\pi^2 k_B^2 T}{3} \right) = \frac{Q}{2 \pi^2} \left(\frac{\pi^2 k_B^2 T}{3} \right),
\eeq
which, alternatively, may also be viewed as a manifestation of the chiral-gravitational mixed anomaly~\cite{Sachdev16,grav_anomaly}. 
In the Supplemental Material we discuss a more formal, but physically equivalent, way to describe the chiral anomaly in a Weyl semimetal~\cite{Note1}.

Tuning the node separation $2 Q$ between $0$ and $2 \pi$ realizes the transition between a trivial and an
integer quantum Hall insulator in 3D~\cite{Halperin_1987,ZhangIQHE3D}, which 
has to proceed through the intermediate Weyl semimetal phase~\cite{Burkov11-1}, unlike in 2D, where there is a critical point (plateau transition). 
The chiral anomaly also leads to the appearance of Fermi arc surface states, since the action in Eq.~\eqref{eq:2} fails to be gauge invariant 
in the presence of a boundary, which makes the existence of a boundary-localized state necessary~\cite{Goswami_anomalies}. 

Apart from giving rise to topological response and protected surface states, anomalies can also place strong restrictions 
on the possible effect of electron-electron interactions.
In particular, anomalies prohibit opening a gap without either breaking the protecting symmetry or creating an exotic state with topological 
order, as was recently discussed extensively in the context of strongly-interacting 2D surface states of 3D symmetry-protected topological orders~\cite{Wen_SPT,Wen_SPT1} in bosonic~\cite{AshvinSenthil} and fermionic~\cite{FidkowskiSurface,Wang13,Metlitski15,Fidkowski14,Bonderson13,WangSenthil14,MetlitskiVortex,PotterWangMetlitskiVishwanath} systems. 
In this Letter, we aim to answer analogous questions in the case of a 3D Weyl semimetal: can one open a gap in a Weyl semimetal without breaking translational or charge conservation symmetries while preserving the chiral and the gravitational anomalies, which lead to the electrical and thermal Hall conductivities of 
Eqs.~\eqref{eq:3} and \eqref{eq:4}? 
What would be the universal properties of such gapped phases?

To answer these questions we will adopt the strategy known as ``vortex condensation", which has been successful in the context of 2D surface states of 3D bulk TI~\cite{Wang13,Metlitski15}. We will start by inducing a phase-coherent superconducting state in a magnetic Weyl semimetal (with only a single pair of nodes for simplicity, although the results readily generalize to any odd number of node pairs), which violates the charge conservation. 
We then attempt to produce a gapped insulator by proliferating vortices and restoring the charge conservation symmetry, while keeping the pairing gap intact. 
In order to make this procedure well-defined, we will assume the superconducting pairing to be weak, i.e. the induced gap is taken to be much smaller than $v_F Q$, 
where $v_F$ is the Fermi velocity of the Weyl cones. In this case it is impossible to gap out the Weyl nodes by simply pushing them to the edge or the center of the BZ, where 
they can mutually annihilate without breaking translational symmetry. In the language of the anomaly, we are demanding that the coefficient of the anomaly $Q$, which takes continuous values and is thus not strictly protected, is fixed throughout the procedure.

It is easy to see that, in this situation, a BCS-type pairing of time-reversed states can not produce a gapped superconductor~\cite{Meng12,Moore12,Bednik15,YiLi18}. 
It is, however, possible to open a gap by inducing a Fulde-Ferrell-Larkin-Ovchinnikov (FFLO)-type superconducting state instead, where states on each side 
of the two Weyl nodes are paired~\cite{Moore12,Bednik15}. 
Since pairing in the FFLO state may (approximately) be taken to occur independently in each Weyl cone, let us consider a single (right-handed) Weyl fermion with singlet pairing
\beq
\label{eq:7}
H = v_F \sum_\bk c^\dg_\bk \bsigma \cdot \bk \,c^\pdg_\bk + \Delta \sum_\bk (c^\dg_{\bk \upa} c^\dg_{- \bk \da} + c^\pdg_{-\bk \da} c^\pdg_{\bk \upa}), 
\eeq
Introducing Nambu spinor $\psi^\pdg_\bk = (c^\pdg_{\bk \upa}, c^\pdg_{\bk \da}, c^\dg_{-\bk \da}, -c^\dg_{-\bk \upa})$, this may be written as
\beq
\label{eq:8}
H = \frac{1}{2} \sum_\bk \psi^\dg_\bk (v_F \tau^z \bsigma \cdot \bk + \Delta \tau^x) \psi^\pdg_\bk,
\eeq
which is simply the Hamiltonian of a Dirac fermion of mass $\Delta$. 
This, however, leads to a density modulation and thus broken translational symmetry. 
Since $\Delta(\bQ) \sim \sum_\bk \langle c^\dg_{\bQ + \bk} c^\dg_{\bQ - \bk} \rangle$ carries momentum $2 \bQ$, a gauge-invariant 
density modulation $\varrho(\bQ) \sim \Delta^*(- \bQ) \Delta(\bQ)$ will carry momentum $4 \bQ$. 
In general, this breaks translational symmetry, which may not be restored even when the superconductivity is destroyed by proliferating 
vortices. 
This is true, except when $\bQ = \bG/4$, where $\bG$ is the smallest nonzero reciprocal lattice vector. 
In this case a gapped FFLO state does not break translational symmetry. 
We will thus concentrate on the $\bQ = \bG/4$ case henceforth. 

An important question is what happens to the Fermi arc surface modes of the Weyl semimetal in the FFLO state. 
The Fermi arc is in principle unaffected by pairing since it is spin-polarized. 
However, due to the effective doubling of degrees of freedom, induced by pairing, which is corrected by the factor of $1/2$ in Eq.~\eqref{eq:8}, the Fermi arc get copied 
to the part of the BZ outside of the Weyl points, and occupies the range of $4 \bQ$, which always coincides with the size of the new BZ, reduced by the translational
symmetry breaking in the FFLO state~\cite{Ting17}. 
When $\bQ = \bG/4$, however, this range is identical to the size of the original BZ, which is another way to see why 
the FFLO state does not break translational symmetry when and only when the Weyl node separation is exactly half the size of the BZ~\footnote{See Supplemental Material for an alternative chiral anomaly formulation, the calculation of the Fermi arc surface state in the FFLO superconductor, the derivation of the straight-line vortex Majorana modes, and some formal details on vortex condensation}. 
This implies that, while the electrical Hall conductivity in the FFLO state is no longer the same as in the non-superconducting Weyl semimetal due to the breaking of the 
charge conservation symmetry, the thermal Hall conductivity remains unaffected and is determined by the length of the Fermi (Majorana) arc
\beq
\label{eq:9}
\kappa_{xy} = \frac{Q}{2 \pi^2} \left(\frac{\pi^2 k_B^2 T}{3} \right) = \frac{1}{4 \pi} \left(\frac{\pi^2 k_B^2 T}{3} \right).
\eeq
In other words, the chiral-gravitational mixed anomaly is unaffected by the formation of the FFLO state. 

\begin{figure}[t]
\subfigure{
\label{fig:1a}
\includegraphics[width=4cm]{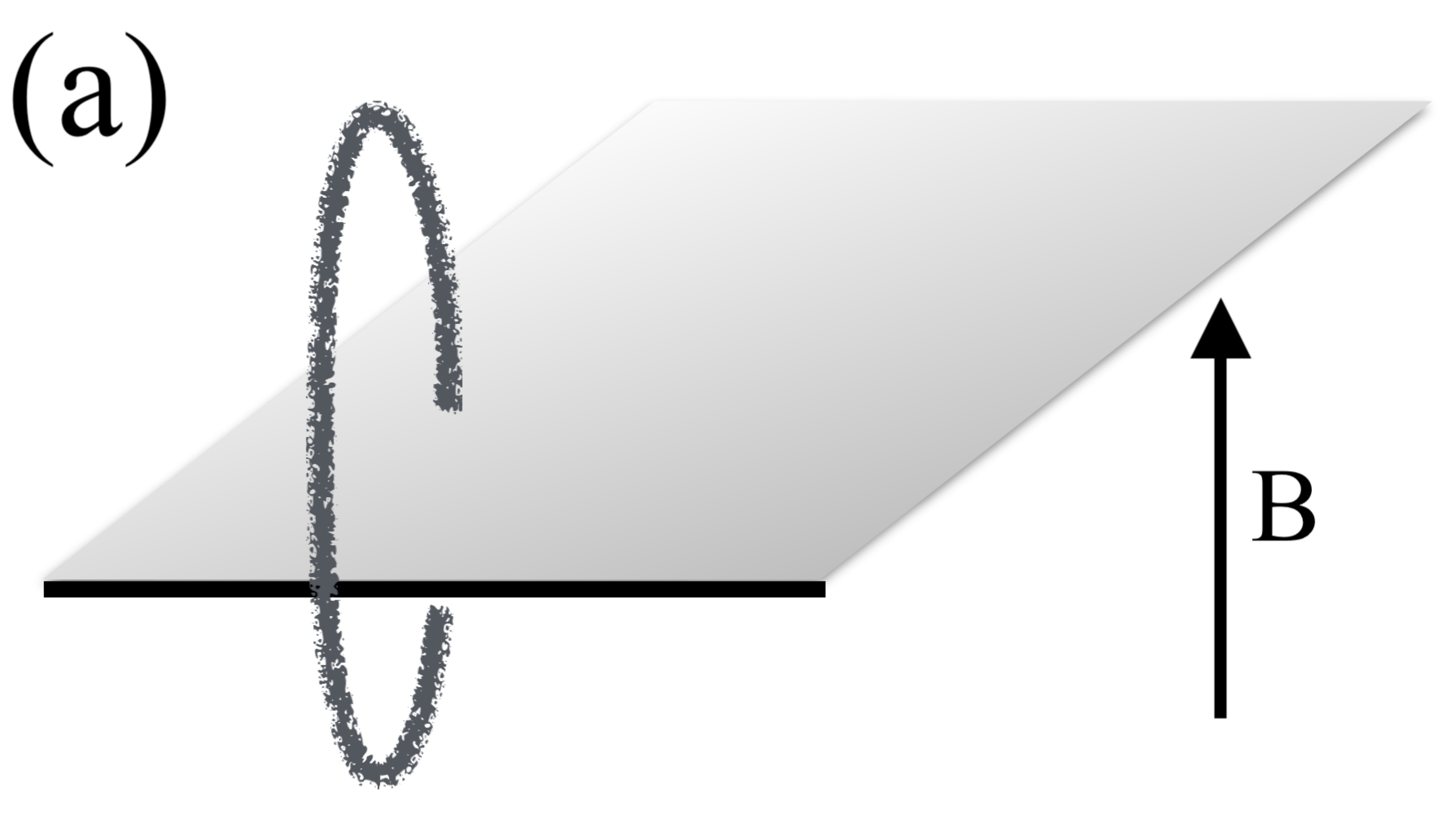}}
\subfigure{
\label{fig:1b}
\includegraphics[width=4cm]{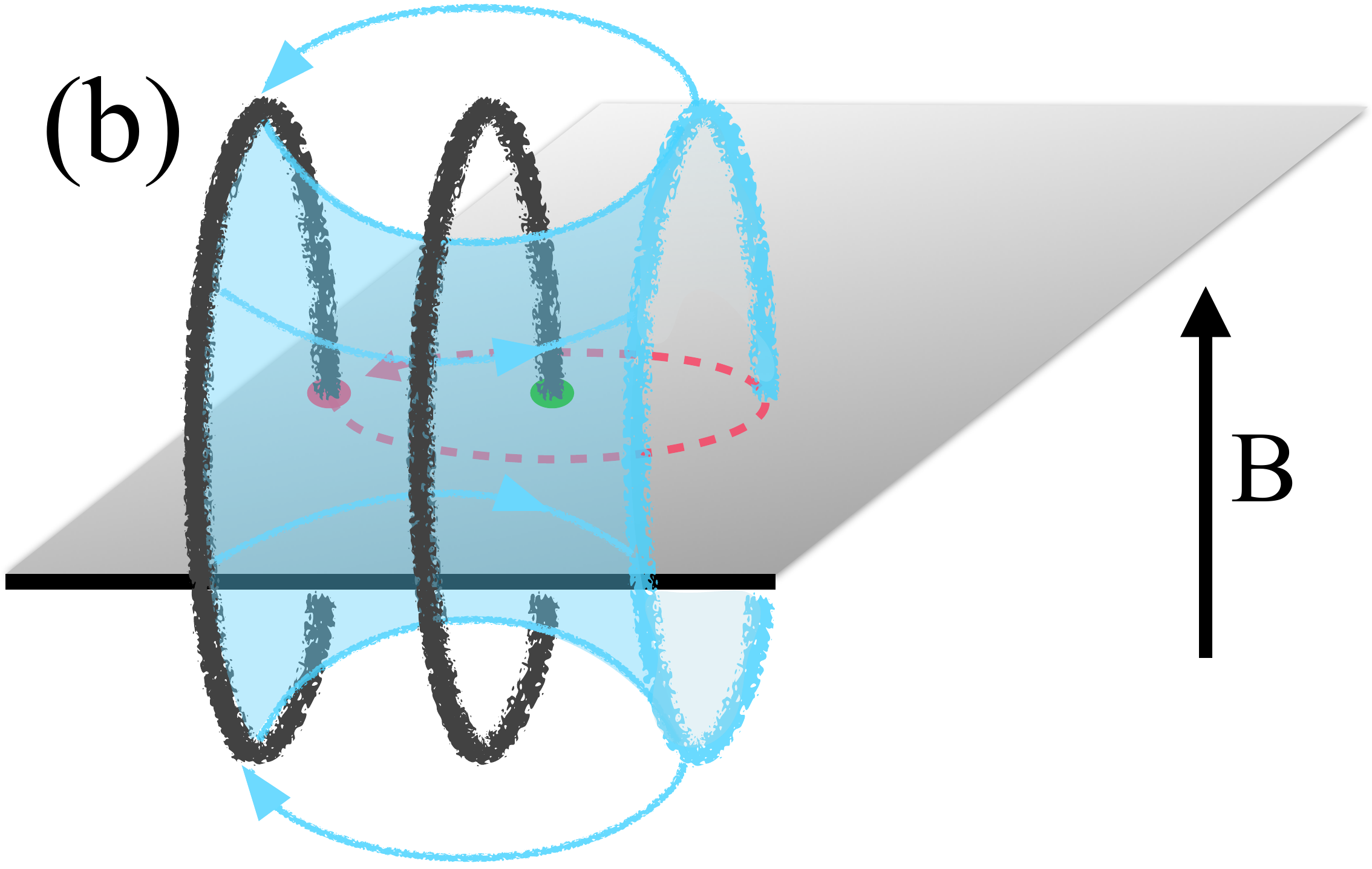}}
\caption{(Color online) (a) A vortex loop linked with a dislocation with the Burgers vector $\bB= \hat z$. Fractional quantum numbers and nontrivial braiding statistics can emerge in such a configuration. (b) A pair of vortex loops linked with a dislocation with the Burgers vector ${\bf B} = \hat z$. Braiding the two loops may be accomplished by adiabatically 
shrinking the left loop, then moving it to the right by crossing the disc, enclosed by the right loop, then expanding and moving it back to the original place without crossing 
the disc, enclosed by the second loop.}
\end{figure}

We now try to restore the charge conservation symmetry by proliferating vortices in the superconducting order parameter while keeping the pairing gap for the Weyl fermions.
If the vortices can be condensed without breaking the translational symmetry, we will obtain a gapped state that is fully symmetric.
This state must have $\sigma_{xy} = 1/4 \pi$ to match the chiral anomaly.
To accomplish this, we need to understand carefully what does it mean to condense vortices, which form loops in 3D, without breaking the translational symmetry. In the simpler case of condensing particles, we would want the particle to carry zero momentum (up to a gauge choice). Now we want to achieve the same goal for vortex loops, which means that we want to condense vortex loops that transform trivially under translation. A good way to probe the properties of a loop under translation is to link the loop to a lattice dislocation with the Burgers vector $\bB=\hat{z}$, which inserts a half $xy$-plane, ending on a dislocation line, as shown in Fig.~\ref{fig:1a}. 
If a vortex is truly trivial under translation, such a link should not create any nontrivial effect. 

Consider first a vortex loop with an odd vorticity, trapping a magnetic flux $\Phi = (2n+1)\pi$. A straightforward calculation shows that each time the vortex penetrates an atomic $xy$-plane, a Majorana zero mode is trapped at the intersection~\cite{Note1}. An ordinary closed loop contains an even number of such zero modes since the $xy$-plane is penetrated an even number of times. However when linked with a dislocation with $\bB = \hat{z}$, the total number of such penetrations becomes odd and the vortex now carries an unpaired Majorana zero mode. 

The effect becomes more drastic when two vortices with an odd vorticity are simultaneously linked to a $\bB = \hat{z}$ dislocation.
In this configuration we can consider braiding between the two vortices, as illustrated in Fig.~\ref{fig:1b}. This process was first discussed in Ref.~\cite{Wang-Levin} and is known as three-loop braiding -- the only difference in our case is that the ``base" loop is a static dislocation rather than a dynamical excitation. Because of the Majorana zero modes, carried by the vortices when linked with the dislocation, the loop braiding process is non-abelian.

The above reasoning shows that odd vortices should be considered nontrivial under translation symmetry and cannot be condensed without breaking the symmetry. Yet another way to see this is that if we were to condense such vortices, inserting a dislocation into the system would require the inserted half-plane to be out of the bulk ground state to cancel the nontrivial braiding statistics of the linked vortices (only then a condensate is possible). This implies an energy cost $\sim O(L^2)$ instead of $\sim O(L)$ for an ordinary dislocation, where $L$ is the system size. This simply means that the translation symmetry has actually been broken in the process. 

Now what about vortices with even vorticity? There is no unpaired Majorana zero mode in this case, even when linked with a 
dislocation~\cite{Note1}. But the braiding statistics between two such vortices, linked with the same dislocation, can still be nontrivial (though must be abelian). Since to match the chiral anomaly we need the Hall conductivity of $\sigma_{xy}=1/4 \pi$ per layer, a two-fold vortex (with flux $\Phi = 2\pi$) will induce a semionic particle with the self-statistical phase $\theta = \pi \sigma_{xy}/(1/2\pi) = \pi/2$ each time it penetrates the $xy$-plane.
As before, an ordinary two-fold vortex loop will not possess nontrivial self-statistics since the $xy$-plane is penetrated twice. But when linked with a $\bB = \hat{z}$ dislocation, each vortex traps an unpaired semion, which leads to semion braiding statistics for the two-loop braiding process in Fig.~\ref{fig:1b}. This nontrivial abelian braiding of $2\pi$ vortices, linked to dislocations, is the fingerprint of the chiral anomaly when the $U(1)$ symmetry is broken. We thus come to the conclusion that two-fold vortices are also nontrivial under translations and cannot be condensed. 

Analogous considerations imply that four-fold ($\Phi = 4\pi$) vortex loops have bosonic statistics even when linked with dislocations and thus may be condensed. 
This produces an insulating state, which does not break either the charge conservation or the translational symmetry and has 
an electrical Hall conductivity $\sigma_{xy} = 1/4 \pi$ and a thermal Hall conductivity $\kappa_{xy} = (1/ 4 \pi) (\pi^2 k_B^2 T/3)$ per layer. 
This is an insulating state that preserves all the symmetries and both the chiral and the gravitational anomaly of a Weyl semimetal with $2 Q  = \pi$.

The insulator thus obtained is not a trivial one -- it possesses a $\mathbb{Z}_4$ topological order~\cite{Balents99,SenthilFisher,Note1}. The uncondensed one-, two- and three-fold vortices survive as nontrivial gapped loop excitations in the topological order, with inherited nontrivial braiding statistics when linked with dislocations. There are also nontrivial particle excitations. The Bogoliubov fermion in the paired state survives as a neutral fermion excitation. The condensation of $4\pi$ vortices also leads to the emergence of a charge-$1/2$ boson as a gapped excitation -- this can be understood as a point defect which, when taken around the condensed $4\pi$ vortex loop, acquires a Berry phase of $2\pi$. Furthermore, due to a nontrivial mutual braiding statistical phase of $\pi$ between a $\pi$ vortex and a $4\pi$ vortex, when linked with a dislocation, the condensation of $4\pi$ vortices will also bind a $1/4$-charge on a $\pi$ vortex.

In fact, all of the above properties are closely related to the 2D topological order obtained on the surface of an electronic TI through vortex condensation~\cite{Wang13,Metlitski15}. This topological order can be viewed as a Moore-Read Pfaffian state plus a neutral antisemion (with the self-statistics angle $-\pi/2$). The only difference in our case is that some of the ``vortex-like" particles in the topological order show up as links between loop excitations and a dislocation with $\bB = \hat{z}$. 

This motivates the following parton construction of the anomalous topological order~\cite{SeibergWitten16,PotterWangMetlitskiVishwanath}. We decompose the electron 
operator as 
\beq
\label{eq:15}
c = b^2 f,
\eeq
where $b$ is a charge-$1/2$ boson, while $f$ is a neutral fermion. The neutral fermion experiences the same electronic structure as the original Weyl semimetal with $2 Q = \pi$ and the FFLO pairing gap, that does not violate translational symmetry. The neutral Fermi arc surface state then leads to the thermal Hall conductivity $\kappa_{xy} = (1/4 \pi) (\pi^2 k_B^2 T/3)$, which is equivalent to a layered $p+ip$ superconductor~\cite{Read-Green}. The charge-$1/2$ bosons form a layered bosonic integer quantum Hall state~\cite{Lu-Vishwanath,Senthil-Levin}. This state has even integer Hall conductance and zero thermal Hall conductance (more details can be found in Refs.~\cite{Lu-Vishwanath,Senthil-Levin,JainBoson}). In our case the bosonic integer quantum Hall state contributes a Hall conductivity $\sigma_{xy}=2(1/2)^2/2\pi = 1/4\pi$ per layer. This gapped insulating state thus reproduces exactly the chiral and the gravitational anomalies of the Weyl semimetal, while preserving its translational and charge conservation symmetries. 

The parton decomposition of Eq.~\eqref{eq:15} and the mean field states of $b$ and $f$ are invariant under a $\mathbb{Z}_4$ gauge transform $b \rightarrow i^n b, \,\, f \rightarrow (-1)^n f, \,\, n \in {\mathbb Z}_4$, 
which is consistent with the $\mathbb{Z}_4$ topological order. One can check explicitly that the $\mathbb{Z}_4$ gauge flux loops have the same properties with the remnants of the uncondensed vortex loops from the vortex-condensation construction. For example, 
a fundamental ($\Phi = \pi$) vortex is seen by the fermion $f$ as a $\pi$ vortex, and therefore leads to a Majorana zero mode whenever the vortex penetrates the $xy$-plane. The fundamental vortex is also seen by the boson $b$ as a $\pi/2$ vortex, which leads to a fractional charge $q=(\pi/2)\sigma_{xy}/(1/2) = 1/4$ whenever the vortex penetrates the $xy$-plane. The bosonic integer quantum Hall state also leads to a semion whenever a two-fold vortex penetrates the $xy$-plane. Again all these properties are sharply manifested when the vortices are linked with dislocations.
\begin{figure}[t]
\vspace{-0.7cm}
\includegraphics[width=8cm]{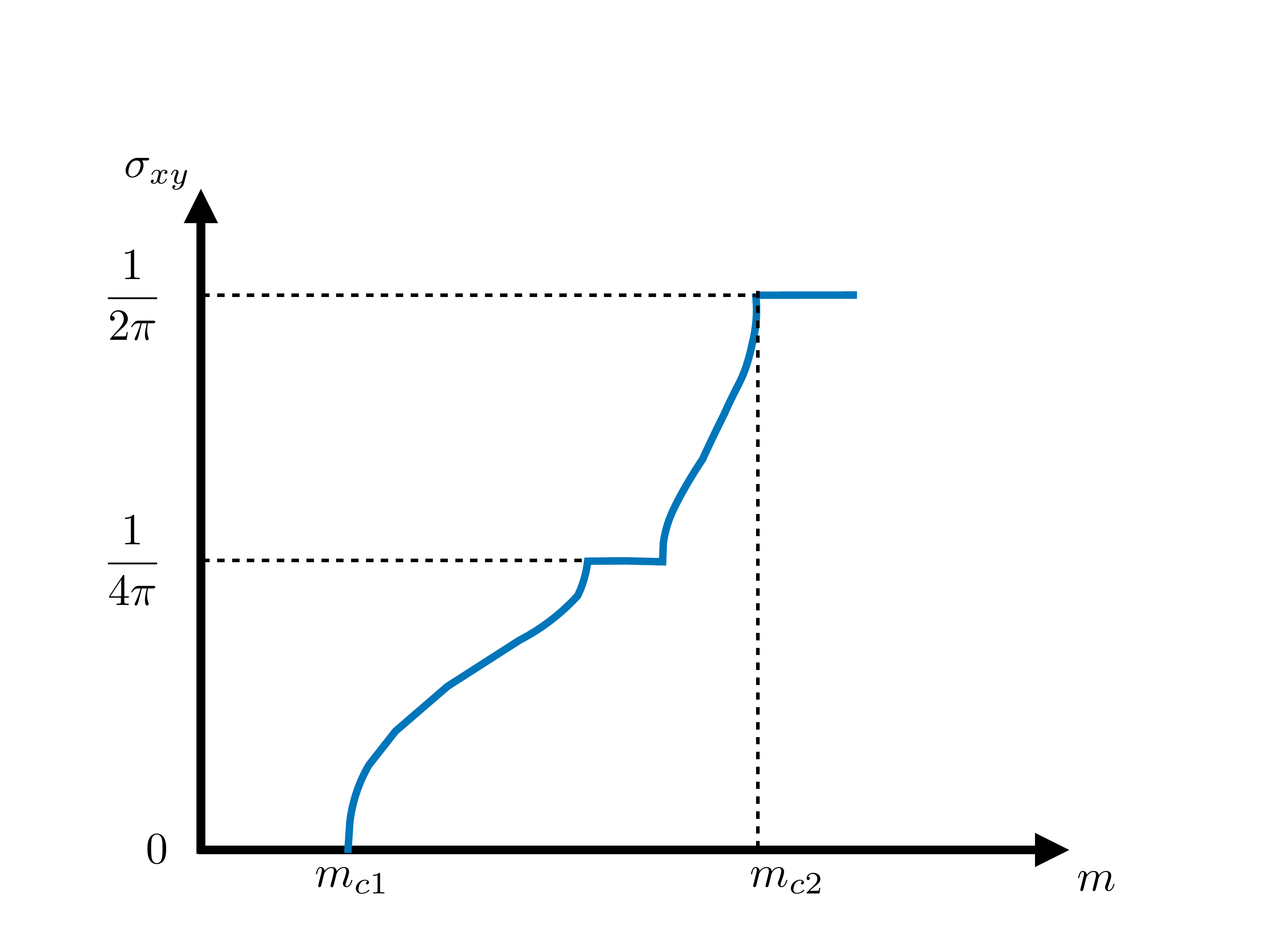}
\caption{(Color online) Hall conductivity as a function of the magnetization with a fractional plateau corresponding to $\sigma_{xy} = 1/ 4 \pi$ 3D FQHE.}
\label{fig:plateau}
\end{figure}

In addition to realizing the chiral and the gravitational anomalies of the Weyl semimetal, the above state also provides a realization of the fractional quantum Hall effect (FQHE) in 3D, which may not be regarded as simple layering of weakly-coupled 2D FQHE systems. 
As discussed above, a magnetic Weyl semimetal with two Weyl nodes is an intermediate phase between an ordinary 3D insulator with 
$\sigma_{xy} = 0$ and an integer quantum Hall insulator with $\sigma_{xy} = 1/2 \pi$. 
We may tune between the two phases by varying a TR-breaking parameter, i.e. magnetization $m$. 
One may view this as an analog of tuning the filling factor by an applied magnetic field in the case of the 2D quantum Hall effect. 
There are two critical values of the magnetization, $m_{c1}$ and $m_{c2}$, which correspond to transitions
from the ordinary insulator to the Weyl semimetal and from the Weyl semimetal to the integer quantum Hall insulator 
correspondingly. 
The function $Q(m)$, which determines the separation between the pair of Weyl points and the Hall conductivity 
$\sigma_{xy}(m) = Q(m)/ 2 \pi^2$ as a function of the 
magnetization, is model-dependent, but becomes universal near each critical point. 
For noninteracting electrons, we have~\cite{Burkov11-1} $Q(m) \sim A_1 (m - m_{c1})^{1/2}, \,\, \pi - A_2 (m_{c2} - m)^{1/2}$,
where $A_{1,2}$ are nonuniversal coefficients. 
We then claim that, in the presence of strong electron-electron interactions, a fractional plateau may exist in $\sigma_{xy}(m)$, at which the Hall conductivity 
is quantized to half the value of the integer plateau, $\sigma_{xy} = 1/ 4 \pi$, as shown in Fig.~\ref{fig:plateau}. 

It is important to note that the constraint on the possible plateau comes mainly from the thermal Hall response. For a topological order that is genuinely three dimensional, in the sense that all excitations can move in all three directions, the particle excitations can only be bosonic or fermionic. This constrains the thermal Hall conductance per layer to be quantized to $\kappa_{xy}=(n/2)(\pi k_B^2T/6)$, where $n$ is odd only if the fermion excitations form layered topological ($p+ip$ - like) superconductors.
Plateaus at other values of $\sigma_{xy}$ are certainly possible, but these states will be unrelated to Weyl semimetals.

In conclusion, in this paper we have addressed the question of whether it is possible to open a gap in a magnetic Weyl semimetal, while not breaking any symmetries and while 
preserving the chiral anomaly, by which we mean the electrical and thermal Hall conductivities, proportional to the Weyl node separation, Eqs.~\eqref{eq:3} and \eqref{eq:4}. When the separation between the Weyl nodes $2 Q$ is not an integer multiple of a primitive reciprocal lattice vector, the resulting ``fractional" electrical (and thermal) Hall 
conductivity prohibits opening a gap in the weakly-interacting regime. 
We have demonstrated that such gap opening is possible, but only when the separation between the Weyl nodes is equal to half a reciprocal lattice vector. 
The state one obtains is a true featureless topologically-ordered 3D liquid and may be viewed as a generalization of the Pfaffian-antisemion state~\cite{FidkowskiSurface,Wang13,Metlitski15,Fidkowski14,Bonderson13,WangSenthil14,MetlitskiVortex,PotterWangMetlitskiVishwanath} of a gapped TI surface to 3D. 
Another fruitful way to think about this state is as a nontrivial (i.e. not related to a stack of weakly-coupled 2D systems) generalization of a FQH liquid. 
As can be seen from our analysis, a general feature of such 3D FQH liquids (with intrinsic 3D topological order) is that there exist loop excitations with nontrivial braiding statistics when linked with lattice dislocations, which is a 3D analog of the nontrivial quasiparticle statistics in 2D FQH liquids (quasiparticles in 3D may only be either bosons or fermions). 
In particular, there is a loop excitation that can be induced by a $2\pi$ magnetic flux loop, with an abelian braiding statistical phase of $4\pi^2 \sigma_{xy}$, when linked with a dislocation with $\bB = \hat{z}$. This is in parallel with the 2D FQHE, where there always exists an anyon (known as ``fluxon") with abelian statistics, determined by the fractional Hall conductance. In addition, out state features loop excitations with nonabelian braiding statistics, when linked with dislocations, along with bosonic and fermionic quasiparticle excitations. 

 We note that effects of strong correlations in topological semimetals have been addressed before in Refs.~\cite{Meng16,Morimoto16,Sagi18,Meng19,Teo19}. Our goal and results differ from these works in several important aspects: (a) we have unambiguously defined the meaning of ``chiral anomaly" in Weyl semimetals through Eq.~\eqref{eq:3} and \eqref{eq:4}; (b) we obtained the universal features of the resulting insulators, namely the low-energy topological excitations such as particles and loops of the resulting topological order; and (c) we assumed spatially local interactions, so that the resulting insulating state (with intrinsic topological orders) cannot be adiabatically connected to any free fermion state.

\begin{acknowledgments}
LG was supported by the Natural Sciences and Engineering Research Council (NSERC) of Canada and by a Vanier Canada Graduate Scholarship. 
AAB was supported by Center for Advancement of Topological Semimetals, an Energy Frontier Research Center funded by the U.S. Department of Energy Office of Science, Office of Basic Energy Sciences, through the Ames Laboratory under
contract DE-AC02-07CH11358. 
Research at Perimeter Institute is supported in part by the Government of Canada through the Department of Innovation, Science and Economic Development and by the Province of Ontario through the Ministry of Economic Development, Job Creation and Trade.
\end{acknowledgments}

\end{document}


\title{Supplemental Material: Fractional Quantum Hall Effect in Weyl Semimetals}
\author{Chong Wang}
\affiliation{Perimeter Institute for Theoretical Physics, Waterloo, Ontario N2L 2Y5, Canada}
\author{L. Gioia}
\affiliation{Department of Physics and Astronomy, University of Waterloo, Waterloo, Ontario 
N2L 3G1, Canada} 
\affiliation{Perimeter Institute for Theoretical Physics, Waterloo, Ontario N2L 2Y5, Canada}
\author{A.A. Burkov}
\affiliation{Department of Physics and Astronomy, University of Waterloo, Waterloo, Ontario 
N2L 3G1, Canada} 
\date{\today}

\maketitle

\section{Another formulation of the chiral anomaly}

Consider a (magnetic) Weyl semimetal with two Weyl nodes in the Brillouin zone separated by a momentum $\Delta \bk = 2Q\hat{z}$. We assume charge neutrality so there is no Fermi surface. The symmetries involved in the ``chiral anomaly" are the $U(1)$ charge conservation and the translation symmetry in $\hat{z}$ (call the group $\mathbb{Z}^z$ and the generator $T_z$).

To see the anomaly, we consider ``gauging" the $U(1)\times\mathbb{Z}^z$ symmetry by introducing probe gauge fields $A_{\mu}$ and $z\in H^1(M,\mathbb{Z})$ ($M$ is the space-time manifold). The integer gauge field $z$ associated with translation symmetry~\cite{ThorngrenElse} has the following properties: it is locally flat ($dz=0$), and the Wilson loop $\int_{C_1}z\in\mathbb{Z}$ over a $1$-cycle $C_1$ essentially counts the number of $\hat{z}$-translations around $C_1$, and a unit defect in $z$ represents a lattice dislocation with Burgers vector $\vec{B}=\hat{z}$. To probe the thermal response, we also couple the system (in the continuum limit) to a metric $g$. The chiral anomaly can be understood as a $(4+1)d$ ``bulk" term:
\beq
\label{ChiralAnomaly}
S_{CA}=i2Q\int_{M_5} z\cup\left(\frac{1}{2}\frac{dA}{2\pi}\wedge\frac{dA}{2\pi}+\frac{1}{192\pi^2}R\wedge R \right),
\eeq
where $R$ is the Riemann curvature. The expression in the parenthesis takes integer value on a $4$-cycle -- this is nothing but the well-known statement that the periodicity of $\Theta$-angle for charged fermions in $(3+1)D$ is $2\pi$ (more formally $A$ is a spin$_c$ connection instead of an ordinary $U(1)$ gauge field). Therefore the coefficient $2Q$ takes continuous value in $[0,2\pi)$, which is consistent with the interpretation that $2Q$ is the momentum separation between the two Weyl nodes in the non-interacting limit. Of course even for interacting fermions the expression Eq.~\eqref{ChiralAnomaly} still makes sense.

The physical interpretation of Eq.~\eqref{ChiralAnomaly} is actually quite simple: there is a Hall conductance per layer $\sigma_{xy}=\frac{\Delta k}{2\pi}$ and thermal Hall conductance per layer $\kappa_{xy}=\frac{\Delta k}{2\pi} (\pi^2k_B^2T/3)$, as already discussed in the main text. For $2Q=2\pi$, these conditions can be satisfied by a gapped state which is equivalent to a stack (in $\hat{z}$ direction) of $2D$ integer quantum Hall states, which is why the anomaly disappears. Again this phenomena is well known from band theory, and we emphasize that it remains well-defined in interacting systems.

The chiral anomaly Eq.~\eqref{ChiralAnomaly} has another consequence, namely a $U(1)$ instanton with $\int d^3x dt\, \bE \cdot \bB = 4\pi^2$ carries a lattice momentum $\bk_{ins}=2Q \hat{z}$. In the non-interacting limit this follows from textbook $U(1)\times U(1)$ chiral anomaly, but the statement also makes sense in interacting systems (where the axial $U(1)$ no longer makes sense but the translation $\mathbb{Z}^z$ does). When charged degrees of freedom are gapless (as in the semimetal phase), the lattice momentum overlaps with physical charge current, which leads to a charge current induced by the instanton -- this is the well-known magnetoresistance from Weyl semimetal\cite{Parameswaran14}. If the system becomes gapped but still preserves the chiral anomaly as defined in Eq.~\eqref{ChiralAnomaly}, the instanton will still carry lattice momentum, but may not induce a charge current. 

We also comment that the anomaly Eq.~\eqref{ChiralAnomaly} is different from the ones typically encountered in the physics of symmetry-protected topological phases, in the sense that the coefficient $2Q$ is not quantized, and therefore is not strictly ``protected". Physically this is simply because one can always smoothly bring the two Weyl cones to the same momentum point and eliminate the anomaly. In our study we keep $2Q$ fixed based on the intuition that we demand nontrivial interactions to take effect well below the electron band width. In the most general parameter space this corresponds to fine tuning one parameter. This is similar to another much more familiar situation: in an ordinary metal the charge density is fixed by tuning the chemical potential. There is also an ``anomaly" associated with non-integer charge density, which is related to the celebrated Lieb-Schultz-Mattis theorem~\cite{LSM,Oshikawa,Hasting}. In this case the anomaly also comes with a non-quantized coefficient (the density), and generically require fine tuning the chemical 
potential~\cite{Song19}.

\section{Fermi arcs in the FFLO Weyl superconductor}

In this section we discuss the effects of the intranodal s-wave interaction on the Fermi arcs of a magnetic Weyl semimetal. We demonstrate the existence of the Fermi arc surface states despite the FFLO pairing interactions creating a bulk gap~\cite{Ting17}.

 To study the Fermi arcs, we need to extend the linearized Weyl Hamiltonian, used in the main text, to the whole Brillouin zone. We choose the following regularized Hamiltonian
\begin{align}
H_0=\sin k_x \sigma^x &+ \sin k_y\sigma^y-\left(\cos k_z-\cos Q\right)\sigma^z\nonumber\\&+m\left(2-\cos k_x-\cos k_y\right)\sigma^z\quad,
\label{eq:h0}
\end{align}
where $Q=\pi/2$ and the lattice constant has been set to one. 

\begin{figure}[hbt]
\vspace{0.5cm}
\includegraphics[width=0.8\columnwidth]{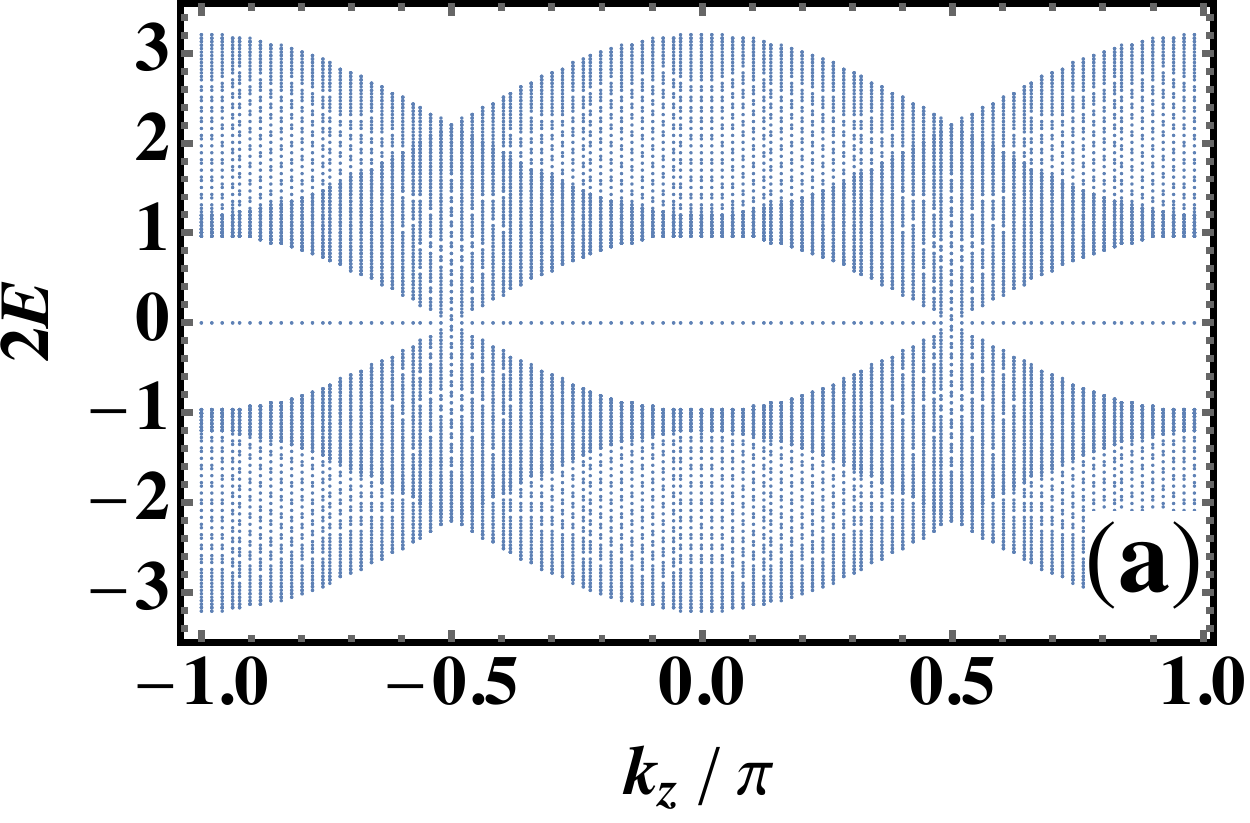}

\includegraphics[width=0.8\columnwidth]{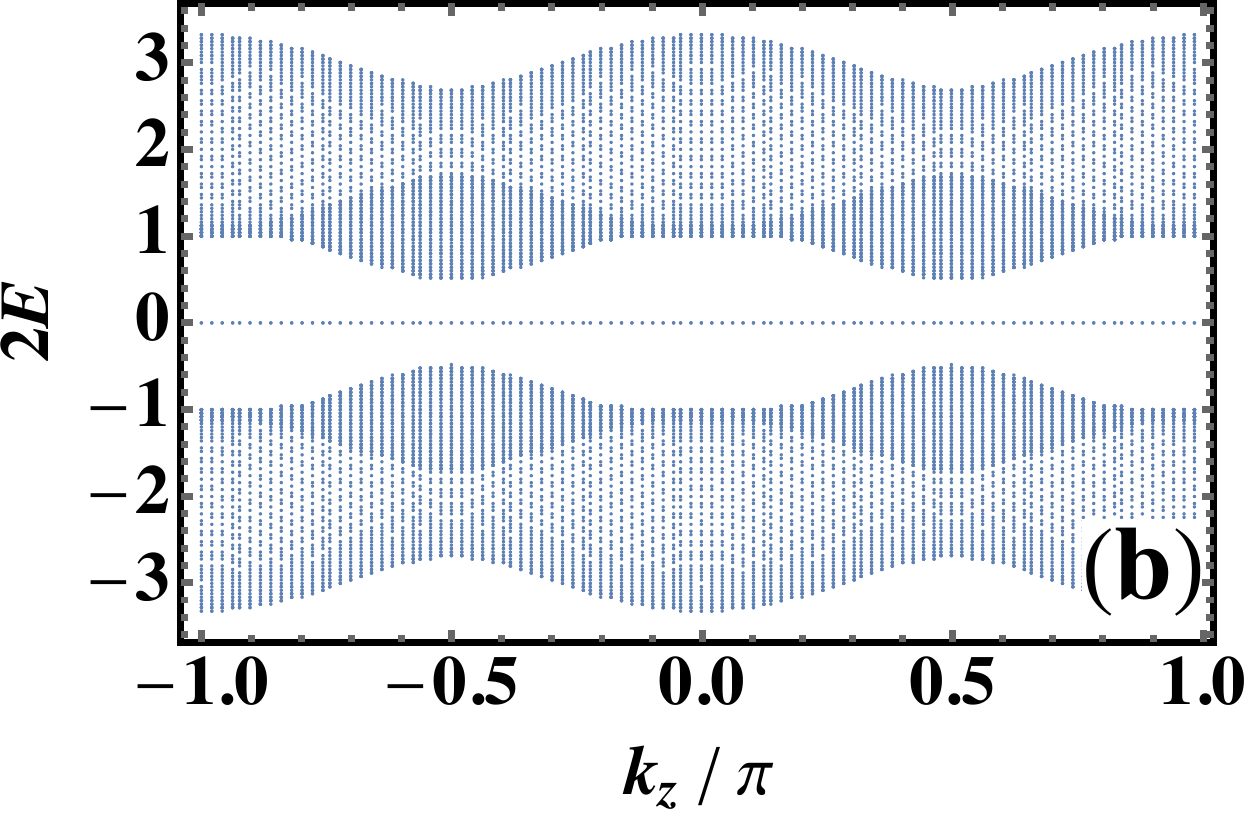}
\caption{\label{fig:Fermiarcnambu} Energy eigenstates for $k_y=0$ slice along $k_z$ corresponding to Nambu Hamiltonians given by Eq.~\ref{eq:h0nambufalt} (a) and \ref{eq:h0int} (b). (a) The zero mode spans the entire Brillouin zone due to doubling of degrees of freedom in the Nambu picture. (b) Intranodal interaction gaps out the bulk Weyl nodes but leaves the surface states unaltered. For both figures $m=1.1$, $N_x=50$ and for (b) $\Delta=0.5$.}
\end{figure}

The superconducting s-wave coupling occurs intranodally, i.e. coupling states with momentum $\mathbf{k}$ to those of momentum $2\mathbf{Q}-\mathbf{k}$, where $\mathbf{Q}=(0,0,\pi/2)$. This situation requires a Nambu basis $\Phi^\dag_\mathbf{k}=(c^\dag_{\mathbf{k}\uparrow},c^\dag_{\mathbf{k}\downarrow},c_{2\mathbf{Q}-\mathbf{k}\uparrow},c_{2\mathbf{Q}-\mathbf{k}\downarrow})$ and the Nambu Hamiltonian then takes the form
\begin{align}
H^{\mathrm{Nambu}}_0=\frac{1}{2}\sum_{\mathbf{k}}\Phi_\mathbf{k}^\dag
\begin{pmatrix}
H_0(\mathbf{k}) & 0 \\
0 & -H_0^\mathrm{T}(2\mathbf{Q}-\mathbf{k})
\end{pmatrix}\Phi_\mathbf{k}\quad.
\label{eq:h0nambualt}
\end{align}
In order to calculate the Fermi arc surface states, we break the translational symmetry along the $x$-direction, leaving a finite-size sample of $N_x$ atomic layers. This leaves $k_y$ and $k_z$ as good quantum numbers. Fourier transforming $k_x$ to a real-space coordinates $n_x$, the Hamiltonian takes the form
\beqa
H^{\mathrm{Nambu}}_0&=&\frac{1}{2}\sum_{n_xk_yk_z}\Phi_{n_xk_yk_z}^\dag \left[h(k_y,k_z) \Phi_{n_xk_yk_z}\right.\nonumber\\
&+&\left.h_+ \Phi_{n_x+1k_yk_z}+ h_- \Phi_{n_x-1k_yk_z}\right].\nonumber \\
\label{eq:h0nambufalt}
\eeqa
with
\begin{align}
h(k_y,k_z)&=
\begin{pmatrix}
h_0(k_y,k_z) & 0 \\
0 & - h_0^\mathrm{T}(-k_y,2Q-k_z)
\end{pmatrix}\quad,\\
h_+&=-\frac{1}{2}\left(i \sigma^x+m\tau^z\sigma^z\right)= h_-^\dag\quad,
\end{align}
where $h_0(k_y,k_z)=\sin k_y \sigma^y+ \big[-(\cos k_z-\cos Q)+m (2-\cos k_y)\big]\sigma^z$, and $\btau$ are Pauli matrices in the Nambu pseudospin space. Diagonalizing this Hamiltonian at every $k_z$ for the $k_y=0$ slice gives eigenenergies as shown in Fig.~\ref{fig:Fermiarcnambu}a. Since the components of the Nambu spinor involve states at momenta $\mathbf{k}$ and $2\mathbf{Q}-\mathbf{k}$, the set of eigenenergies effectively involve states, shifted by $2Q=\pi$ relative to each other. The same principle applies to the Fermi surface states which now span the whole Brillouin zone. Notice that this is purely the result of the doubling of degrees of freedom in the Nambu formalism. Naturally this has no effect on the bulk Weyl nodes at $k_z=\pm Q$.

Now let us observe what happens when we couple the two Nambu copies in Eq.~\ref{eq:h0nambualt} with the FFLO pairing interaction of the form
\begin{align}
H_{\mathrm{int}}=\frac{1}{2}\sum_{\mathbf{k}} \Phi_\mathbf{k}^\dag\Delta \tau^y \sigma^y \Phi_\mathbf{k}\quad.
\end{align}
Under a Fourier transform of $k_x$, we arrive at
\begin{align}
H_{\mathrm{int}}=\frac{\Delta}{2}\sum_{n_xk_yk_z} \Phi_{n_xk_yk_z}^\dag\tau^y\sigma^y \Phi_{n_xk_yk_z}\quad,
\end{align}
such that our full Hamiltonian is now given by
\begin{align}
H=H^{\mathrm{Nambu}}_0+H_{\mathrm{int}}\quad.
\label{eq:h0int}
\end{align}
When this full Hamiltonian is now diagonalised at every $k_z$ for the $k_y=0$ slice we arrive at Fig.~\ref{fig:Fermiarcnambu}b. 
While the bulk states are now gapped, the Fermi arc, spanning the whole Brillouin zone, remains unaffected. 

\section{Helical Majorana fermions in a vertical vortex line}
Let us consider a straight-line vortex of vorticity $n$ along the $z$-direction, which we will take to coincide with the direction of the vector $2 \bQ$, separating 
the pair of Weyl nodes. 
The corresponding Bogoliubov-de Gennes (BdG) Hamiltonian is given by
\beqa
\label{eq:10}
H&=&- i v_F \tau^z (\sigma^x \partial_x + \sigma^y \partial_y) + v_F \tau^z \sigma^z k_z \nonumber \\
&+&\Delta(r) [\cos(n \theta) \tau^x - \sin (n \theta) \tau^y], 
\eeqa
where $r = \sqrt{x^2 + y^2}$, $\Delta(r)$ is the real magnitude of the superconducting order parameter and $\theta$ is the azimuthal angle in the $xy$-plane. 
The eigenstates of $H$ may be easily found explicitly if one assumes $\Delta(r) = \Delta = \textrm{const}$~\cite{Chamon16}. 
In this case one finds exactly $n$ chiral modes, localized in the vortex core, with the following wavefunctions
\beqa
\label{eq:11}
\Psi_{p k_z}(\br) = \frac{(\Delta r / v_F)^{\frac{n}{2}}}{\sqrt{{\cal N}_p}} \left( 
\begin{array}{c}
e^{i \frac{\pi}{4}} e^{i (p - 1) \theta} K_{\frac{n}{2}-p+1}\left(\frac{\Delta r}{v_F}\right)\\  0\\  0\\
e^{-i \frac{\pi}{4}} e^{-i (n - p) \theta} K_{\frac{n}{2}-p}\left(\frac{\Delta r}{v_F}\right)
\end{array}
\right), \nonumber \\
\eeqa
where ${\cal N}_p$ is a normalization factor given by
\beq
\label{eq:12}
{\cal N}_p = \frac{\pi^{3/2} v_F^2}{\Delta^2} \frac{\Gamma(1+n/2) \Gamma(n-p+1)\Gamma(p)}{\Gamma(n/2+1/2)}, 
\eeq
which is finite and positive when $p = 1,\ldots,n$. 
As mentioned above, these localized modes are chiral with the dispersion $\epsilon_p(k_z) = v_F k_z$. 
The degeneracy of the chiral modes with respect to the eigenvalue $p$ is not protected and is lifted when perturbations, such as a finite Fermi energy, 
are introduced. 
A finite Fermi energy leads to a term $-\epsilon_F \tau^z$ in the BdG Hamiltonian Eq.~\eqref{eq:10}.
The problem may no longer be solved exactly (except at $k_z = 0$), but may be solved perturbatively. 
At first order one obtains
\beq
\label{eq:13}
\epsilon_p(k_z)= \epsilon_F\left(1 - \frac{2 p}{n+1} \right) + v_F k_z. 
\eeq
Thus, even though the degeneracy is lifted, exactly $n$ fermionic modes are still always present at zero energy 
in the core of an $n$-fold vortex. 
This is in contrast to the analogous problem of vortex bound states in a superconducting 2D Dirac fermion~\cite{Fukui10},  in which case there is always a single zero mode for odd vorticities and no zero modes for even vorticities. 
The left-handed Weyl node will have an identical set of modes, but with the left-handed dispersion (simply send $v_F\rightarrow-v_F$). 

The nontrivial helical Majorana modes in a straight, vertical vortex with odd vorticity can also be understood using the argument in the main text. Recall that a Majorana zero mode is induced whenever a odd vortex penetrates the $xy$-plane. For a straight, vertical vortex the translation $T_z$ is a good symmetry. We can therefore view the vortex line as a 1D translationally-invariant chain with one Majorana zero mode per unit cell. Such a system has a Lieb-Schultz-Mattis type of constraints on the low energy theory, and cannot be gapped without breaking translation symmetry~\cite{TimSUSY}.

For even vorticity, taking $\epsilon_F = 0$, pairs of Majorana modes may be combined into chiral 1D Weyl modes. 
Since the charge conservation is already violated, pairing terms are always present for these Weyl modes, and they are gapped out 
by the ordinary BCS pairing interaction of the form
\beq
\label{eq:13.5}
H  = v_F \sum_{k_z} [k_z c^\dg_{k_z} \tau^z c^\pdg_{k_z} + \Delta (c^\dg_{k_z} i \tau^y c^\dg_{- k_z} + \textrm{h.c.})/2], 
\eeq 
where the eigenvalues of $\tau^z$ label the chirality of the 1D Weyl modes. 
This state is also stable to small fluctuations in $\epsilon_F$ since it is gapped. 
Thus vortices with even vorticity do not have zero modes in their cores. 

\section{Some formal details on vortex condensation in $(3+1)$ dimensions}

Here we briefly review some formal aspects of vortex condensation in $(3+1)$ dimensions. The logic is in fact very similar to that used in $(2+1)$ dimensional vortex condensation.

Consider a charged system in $(3+1)$ dimensions (Euclidean for simplicity), with conserved $U(1)$ current satisfying continuity equation $\partial_{\mu}j_{\mu}=0$. This equation can be solved by re-writing 
\begin{equation}
j_{\mu}=\frac{1}{2\pi}\epsilon_{\mu\nu\lambda\rho}\partial_{\nu}b_{\lambda\rho},
\end{equation}
where $b_{\lambda\rho}$ is an anti-symmetric two-form gauge field, and the normalization is chosen so that a $2\pi$ flux in $\int db$ corresponds to a unit charge of $\int_{space}j^0$ -- namely $b$ obeys standard Dirac quantization. For the rest of this section we take unit charge to be $2e$ (Cooper pair), so the coupling to electromagnetism is $2A_{\mu}j_{\mu}$.

A free Maxwell-like theory of $b$, $\mathcal{L}\sim (\partial_{[\nu}b_{\lambda\rho]})^2$ ($[,]$ represents anti-symmetrization) corresponds to a superconductor. This can be most easily seen by integrating out the $b$ fields (which can be done since the theory is Gaussian), and obtain an effective response theory $\sim A^2$. Vorticity in this superconductor is represented as a two-form antisymmetric ``current" $J_{\mu\nu}$ that couples to the $b$ field through $J_{\mu\nu}b_{\mu\nu}$. Gauge invariance in $b$ (or simply the fact that vortex lines do not terminate) requires a continuity equation on $J$ which reads $\partial_{\mu}J_{\mu\nu}=0$. This can be solved by re-writing
\begin{equation}
\label{aforvortex}
J_{\mu\nu}=\frac{1}{2\pi}(\partial_{\mu}a_{\nu}-\partial_{\nu}a_{\mu}),
\end{equation}
where $a_{\mu}$ is a dynamical $U(1)$ gauge field -- the normalization is chosen so that a $2\pi$ flux loop in $a$ corresponds to a single vortex.

Now a ``vortex condensation" of the simplest kind, where single vortices (together with all higher vortices) have condensed, means that the gauge field $a$ has only a Maxwell action $(\partial_{[\mu}a_{\nu]})^2$. At low energy the Maxwell terms for both $a$ and $b$ becomes irrelevant and we are left with the topological action
\begin{equation}
    \mathcal{L}=\frac{1}{2\pi}\epsilon_{\mu\nu\lambda\rho}b_{\mu\nu}\partial_{\lambda}a_{\rho}-\frac{2}{2\pi}\epsilon_{\mu\nu\lambda\rho}b_{\mu\nu}\partial_{\lambda}A_{\rho}.
\end{equation}
This is also known as a $BF$ theory and describes a gapped phase of matter. With the coefficients in the above equation, this particular $BF$ theory describes a trivial insulator with no intrinsic topological order.

Now consider condensing $n$-fold vortices ($n>1$), leaving all the lower vortices un-condensed. This is formally implemented by writing $a=n\tilde{a}$ in Eq.~\eqref{aforvortex} where $\tilde{a}$ obeys standard Dirac quantization, and introduce Maxwell term for $\tilde{a}$. The physical meaning is that $J$ can only fluctuate in units of $n$, which is what we mean by $n$-fold vortices. Now the resulting BF theory at low energy becomes
\begin{equation}
    \mathcal{L}=\frac{n}{2\pi}\epsilon_{\mu\nu\lambda\rho}b_{\mu\nu}\partial_{\lambda}\tilde{a}_{\rho}-\frac{2}{2\pi}\epsilon_{\mu\nu\lambda\rho}b_{\mu\nu}\partial_{\lambda}A_{\rho}.
\end{equation}
This is known to represent a $\mathbb{Z}_n$ topological order, with particle charges of $\tilde{a}$ being the topological particle excitations, and line charges of $b$ being the topological loop excitations. Charge fractionalization on the particles can also be seen by introducing a $\tilde{j}_{\mu}\tilde{a}_{\mu}$ term and taking variation on $b$, which leads to $\frac{2}{n}A_{\mu}\tilde{j}_{\mu}$, meaning that the topological quasi-particle carries electric charge $2/n$. The example considered in this work corresponds to $n=4$.